# Universal framework for the design of near-zero refractive index photonic crystals

Adrien Debacq[1], Michaël Lobet[1*]

[1] Department of Physics and Namur Institute of Structured Materials, University of Namur, Rue de Bruxelles 61, 5000 Namur, Belgium

**Emails of the authors:**

Adrien Debacq: adrien.debacq@unamur.be
Michaël Lobet: michael.lobet@unamur.be

*Corresponding Author

# Abstract

Near-zero-index (NZI) media are of great interest for controlling light–matter interactions. However, homogeneous NZI materials operating in the telecom band remain limited and generally suffer from significant absorption losses. In this context, photonic crystals have emerged as an attractive alternative, offering frequency tunability together with the use of low-loss constituent materials. This work develops a predictive method for the design of any of the three classes of NZI, i.e. epsilon-near-zero (ENZ), mu-near-zero (MNZ), or epsilon-and-mu-near-zero (EMNZ) photonic crystals. We present a general demonstration of the equivalence between Dirac cones and EMNZ media, showing that the presence of a Dirac cone at the Γ-point is a necessary and sufficient condition to lead to an effective EMNZ response under normal incidence and that no other dispersion relation can reproduce this behavior in this regime. Consequently, predicting the occurrence of a Dirac cone at the Γ-point enables control of the NZI response of the photonic crystal, whatever the constituent material. This prediction is achieved using the universality of mode symmetries governing the formation of Dirac cones, together with effective medium considerations. Furthermore, the proposed approach allows controlled switching of the effective response between ENZ–MNZ and EMNZ regimes through geometric tuning of the unit cell. This methodology has been successfully applied to both triangular and rectangular lattice configurations, demonstrating its generality across different photonic crystal geometries. These findings establish a predictive framework for the design of NZI photonic crystals and provide an alternative to empirical trial and error optimization strategies. Beyond design applications, they may have implications for the development of metamaterials operating in the telecom regime, as well as for emerging technologies in quantum communication.

**Keywords:** near-zero index materials, photonic crystals, Dirac cone, epsilon near-zero, mu near-zero, epsilon-and-mu near-zero

**TOC Graphic:**

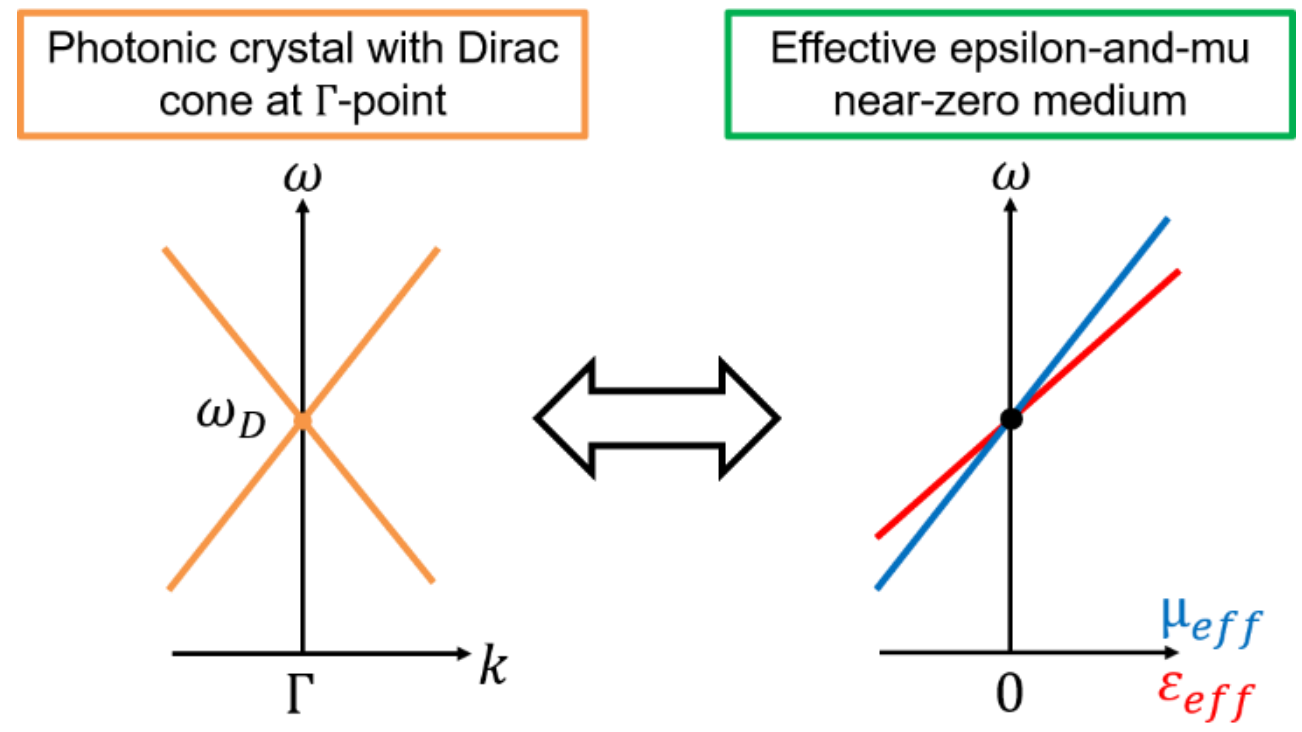

# 1 Introduction

Materials with a refractive index close to zero have attracted considerable attention due to their unique electromagnetic properties, such as an effectively infinite wavelength and a nearly uniform phase throughout the material[1–6]. These features enable a wide range of novel optical phenomena and applications, including cloaking[7,8], supercoupling[9], tailoring radiation phase pattern[10], long-range collective interactions and superradiance[11–17] in quantum optics, and enhanced nonlinear optical processes[18–21] resulting from strengthened light–matter interactions.

Near-zero-index (NZI) materials can be divided into three main categories depending on which constitutive parameter approaches zero: Epsilon-Near-Zero (ENZ) materials, for which the electric permittivity $\varepsilon$ is close to zero; Mu-Near-Zero (MNZ) materials, characterized by a near-zero magnetic permeability $\mu$ and Epsilon-and-Mu-Near-Zero (EMNZ) materials, in which both parameters simultaneously approach zero[1–5]. Although all three classes exhibit a refractive index $n = \sqrt{\varepsilon\mu}$ close to zero, their electromagnetic properties can differ significantly. In particular, the wave impedance $Z = \sqrt{\mu/\varepsilon}$ and group index $n_g$ strongly depend on the specific NZI class. While ENZ and MNZ media exhibit respectively infinite and vanishing impedances, their group indices diverge in the ideal lossless limit[22–24]. In contrast, EMNZ materials can simultaneously support finite impedance and finite, non-zero group index values[4,25–28]. These differences have important practical consequences. For instance, the extreme impedance mismatch associated with ENZ and MNZ media can hinder efficient coupling of electromagnetic waves into the material[2]. The nature of the supercoupling effect, i.e. transmission without phase alteration, also depends on the NZI class: ENZ media exhibit supercoupling through narrow channels[9,29,30], MNZ media through wide channels[31], while EMNZ media allow transmission through arbitrarily shaped channels[8,27,28,32]. Moreover, the spontaneous emission properties of embedded emitters are highly sensitive to both the NZI class and the dimensionality of the system[23]. For example, the Purcell factor diverges in 1D ENZ structure[33–35] but vanishes in 3D EMNZ[32]. Similarly, long-range entanglement and superradiant phenomena are class dependent[11–13]. Consequently, the ability to engineer and realize different classes of NZI materials has become a key objective for tailoring light–matter interactions and exploiting the unique functionalities of near-zero-index systems.

Several material platforms have been developed to realize near-zero-index behavior[1–6]. Natural ENZ responses can be found in materials whose permittivity crosses zero at specific frequencies, typically near a plasma resonance. However, taking the telecommunication band as an example, the range of available materials remains limited and is often associated with significant absorption losses[1]. Besides bulk materials, ENZ behavior can also be achieved in plasmonic waveguides operated close to their cutoff frequency[14,15,33–37]. Nevertheless, losses often limit the performance of these approaches[33,34]. In contrast, naturally occurring MNZ materials are essentially absent at optical frequencies. As a result, MNZ media are predominantly realized using metamaterials composed of subwavelength resonant elements engineered to produce an effective magnetic response. Typical implementations rely on structures such as double split-ring resonators[38] or arrays of metallic inclusions[39], which can be designed to drive the effective permeability towards zero over a narrow frequency range. Nevertheless, low-loss road towards MNZ property is under investigated. Similarly, EMNZ media do not occur naturally and require artificial engineering of both the effective permittivity and permeability. One possible approach to reach the EMNZ class consists of introducing dielectric inclusions into an ENZ host medium[40]. Although several metamaterial-based implementations have been proposed[41,42], photonic crystals have emerged as one of the most attractive platforms for realizing EMNZ behavior[3–6]. Owing to the scalability of Maxwell's equations, photonic crystals provide considerable flexibility in the choice of operating wavelength while enabling the use of low-loss dielectric materials. Consequently, they have become a widely adopted route for achieving near-zero-index phenomena with reduced absorption losses.

Therefore, the realization of EMNZ media using photonic crystals has attracted considerable focus over the past decade[3,4,6,43]. Several studies have demonstrated that photonic crystals exhibiting a Dirac cone at the $\Gamma$-point can display near-zero-index behavior[6,8,43]. In photonics, Dirac-like cones refer to linear band crossings, by analogy with the electronic band structure of graphene[44,45], and typically arise from accidental degeneracies of three modes, where two bands form a linear Dirac cone while the third exhibits an almost flat, quadratic dispersion[4,6]. Such Dirac-like dispersions have been reported and associated to EMNZ behavior in square[8,26–28,46], triangular[27,47,48], and rectangular[49,50] lattices, suggesting that the phenomenon is not restricted to a specific crystal symmetry. As a result, the presence of a Dirac cone at the center of the Brillouin zone is now widely regarded as a key signature of EMNZ behavior in photonic crystals.

However, the equivalence between a photonic crystal with a Dirac cone and an ideal EMNZ medium remains subject to several limitations[43,51–53]. A first issue arises under oblique incidence. While a photonic crystal may behave as a zero-index medium around the Dirac frequency for normally incident waves, additional Bloch modes can be excited at non-zero incidence angles, particularly when flat bands are present nearby[54]. These extra propagation channels allow light transmission under conditions where an ideal isotropic homogeneous zero-index medium would instead reflect the incoming wave[3,51,52], indicating that the zero-index analogy is only strictly valid near normal incidence. A second limitation stems from effective medium theory[55–57]. Several works have suggested that a photonic crystal can only be mapped onto an EMNZ medium when the Dirac cone originates from an accidental degeneracy between monopole and dipole modes at the $\Gamma$-point[6,43,53] and occurs in the long-wavelength regime[4,55], where homogenization remains valid. Furthermore, the Dirac cone should constitute the only accessible propagating channel in the frequency range of interest[56,57]. These observations suggest that a Dirac cone at $\Gamma$-point is a necessary but not sufficient condition for EMNZ behavior at all incidences.

The purpose of this work is to develop a universal and predictive framework for the design of NZI photonic crystals operating around the $\Gamma$-point of the Brillouin zone. To this end, we first establish that the emergence of a Dirac cone at the $\Gamma$-point is a necessary and sufficient condition for obtaining an effective EMNZ response under normal-incidence excitation. Building upon this equivalence, we formulate a predictive design methodology in which the search for near-zero-index photonic crystals is reduced as a symmetry problem. By classifying the photonic modes using group theory, the geometrical conditions required for Dirac-cone formation can be systematically identified. We then show how this framework can be extended beyond the EMNZ case to systematically engineer ENZ and MNZ responses by lifting the accidental degeneracy responsible for the Dirac cone. The methodology is applied to photonic crystals belonging to the $C_{2v}$ and $C_{4v}$ symmetry groups, demonstrating that near-zero-index behavior can be predicted directly from symmetry considerations and mode degeneracies whatever the operating frequency and material used. Together, these results provide a unified framework for the analysis and design of EMNZ, ENZ, and MNZ photonic crystals with applications in telecommunications and quantum operations.

# 2 Results and Discussion

## Design of near-zero-index photonic crystals

Our universal framework assumes that the emergence of a Dirac cone at the $\Gamma$-point of a photonic band structure is a sufficient and necessary condition for obtaining a near-zero-index response at normal incidence. Consequently, predicting the presence of a Dirac cone provides a direct route toward the design of EMNZ photonic crystals (Figure 1). Previous work showed that the formation of Dirac cones is governed by the symmetry properties of the photonic modes rather than by the specific geometrical details of a given structure[58,59]. The concept of the universality of mode symmetry for creating Dirac cones establishes a set of general rules linking mode degeneracies at the $\Gamma$-point to the resulting band structure[59]. In particular, it identifies which combinations of modes can produce linear dispersions for $C_{2v}$, $C_{4v}$ and $C_{6v}$ 2D photonic crystals under appropriate conditions[60,61]. The main implication is that the search for near-zero-index photonic crystals can be reformulated as a symmetry problem. Once the symmetry group and material platform are chosen, the design process reduces to identifying the geometrical parameters that bring the relevant modes into accidental degeneracy (Figure 1). In practice, this is achieved by tracking the eigenfrequencies of the modes at the $\Gamma$-point as a function of the geometric parameters of the unit cell. The dimensions corresponding to the crossing of the target modes then directly provide the conditions required for the formation of a Dirac cone. By combining Sakoda's universality rules[58,59] with a classification based on group theory of the photonic modes, this framework predicts whether a given photonic crystal can support a Dirac cone and therefore exhibit EMNZ behavior.

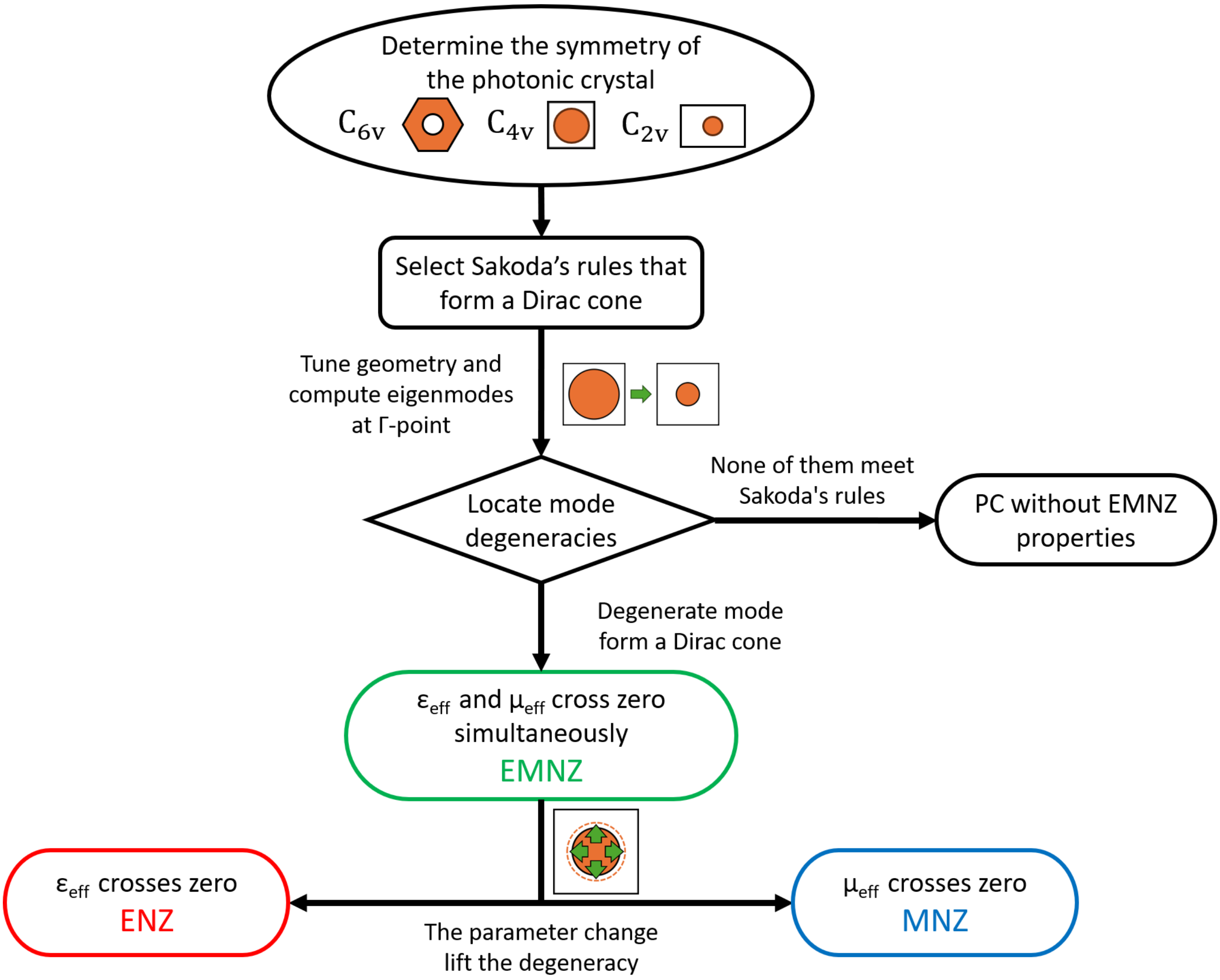

*Figure 1: Flowchart of the predictive framework for the design of near-zero-index photonic crystals. Starting from the lattice symmetry, Sakoda's rules based on group theory identify the mode degeneracies at the Γ-point that can support a Dirac cone. By tuning the geometrical parameters, these degeneracies are achieved, leading to an EMNZ regime where the effective permittivity and permeability simultaneously vanish at the Dirac point. A slight lifting of the degeneracy opens a band gap, resulting in a purely imaginary effective refractive index and regions where permittivity and permeability have opposite signs. This gives rise to ENZ and MNZ responses on either side of the gap, providing a unified route to predict and engineer EMNZ, ENZ, and MNZ behaviors directly from the symmetry and geometry of the photonic crystal.*

# Dirac cone–EMNZ equivalence

Before applying this framework, we first establish the equivalence between Dirac cone formation at the Γ-point and the emergence of an EMNZ response at normal incidence. Previous work established that a Dirac cone located at the Γ-point is associated with an effective EMNZ response[8,62]. The physical origin of this correspondence can be understood from the linear dispersion relation around the Dirac frequency $\omega_D$,

$$\omega(k) = \omega_D + v_g k, \tag{1}$$

where $v_g$ is the group velocity associated with the slope of the cone. Comparing this expression with the dispersion relation of an effective homogeneous medium, $k = n_{eff}\omega/c$, yields to an effective refractive $n_{eff}$ index equals to:

$$n_{eff} = n_g \frac{\omega - \omega_D}{\omega}, \tag{2}$$

with $n_g = c/v_g$ the group index. The effective refractive index therefore vanishes at the Dirac frequency and changes sign across it, exhibiting a characteristic negative-zero-positive behavior[62]. Since such a sign transition requires the effective permittivity and permeability to simultaneously cross zero, the presence of a Dirac cone at the $\Gamma$-point directly implies an EMNZ response (Figure 2a).

We now demonstrate the converse statement, namely that an EMNZ response necessarily implies the existence of a Dirac cone, i.e. linear dispersion. Consider a homogeneous medium for which the effective permittivity and permeability simultaneously vanish at a frequency $\omega_D$ ($\varepsilon(\omega_D) = \mu(\omega_D) = 0$). The dispersion relation can be written as

$$k^2 = \frac{\omega^2}{c^2}\varepsilon(\omega)\mu(\omega). \tag{3}$$

Near $\omega_D$, both constitutive parameters can be expanded to first order,

$$\begin{aligned} \varepsilon(\omega) &\simeq \varepsilon^{(1)}(\omega - \omega_D), \\ \mu(\omega) &\simeq \mu^{(1)}(\omega - \omega_D), \end{aligned} \tag{4}$$

where the first-order derivatives $\varepsilon^{(1)} = \partial\varepsilon/\partial\omega|_{\omega_D}$ and $\mu^{(1)} = \partial\mu/\partial\omega|_{\omega_D}$ are assumed to be finite and strictly positive[63]. Substituting these expressions into the dispersion relation near $\omega_D$ yields

$$k^2 \simeq \frac{\omega_D^2}{c^2}\,\varepsilon^{(1)}\mu^{(1)}(\omega - \omega_D)^2. \tag{5}$$

As a result, the frequency becomes a linear function of the wavevector around $\omega_D$,

$$\omega_{\pm(k)} = \omega_D \pm \frac{1}{\sqrt{A}}|k|, \tag{6}$$

with $A = \frac{\omega_D^2}{c^2}\varepsilon^{(1)}\mu^{(1)}$. The dispersion therefore consists of two linear branches intersecting at the point ($k = 0, \omega_D$), which is precisely the signature of a Dirac cone.

To identify the slope of these branches, the group index can be evaluated from the dispersive permittivity and permeability[64]. Near $\omega_D$, it reduces to

$$n_g \simeq \omega_D\sqrt{\varepsilon^{(1)}\mu^{(1)}}, \tag{7}$$

leading to a group velocity $v_g = c/n_g = A^{-\frac{1}{2}}$. The dispersion relation can therefore be rewritten as

$$\omega_{\pm(k)} = \omega_D \pm v_g|k|, \tag{8}$$

which corresponds exactly to the equation of a Dirac cone (equation 1). Hence, provided that the effective permittivity and permeability simultaneously cross zero with non-vanishing first-order derivatives, an EMNZ medium necessarily exhibits a Dirac-cone dispersion at the zero-index frequency (Figure 2a). It should be noted that the condition on the non-vanishing first-order derivatives of both the permittivity and the permeability is nothing but a requirement that effective NZI property is dispersive, which is directly imposed by causality considerations[4,22,63].

The above established equivalence between Dirac-cone dispersion and EMNZ behavior also provides a direct route toward the design of ENZ and MNZ photonic crystals[4,11,12]. We confirmed that at the Dirac point, the effective permittivity and permeability simultaneously cross zero. When the accidental degeneracy responsible for the Dirac cone is slightly lifted through a modification of the geometrical parameters, a band gap opens around the former Dirac frequency (Figure 2b). Within this gap, the refractive index becomes purely imaginary, indicating that the effective permittivity and permeability have opposite signs. Consequently, their zero-crossings are displaced to opposite sides of the band gap, creating an ENZ region on one side and an MNZ region on the other[4] (Figure 2b).

This observation naturally extends the predictive framework introduced above (Figure 1). The search for ENZ and MNZ photonic crystals starts by identifying the geometrical parameters that produce the target Dirac cone. Small perturbations around this accidental degeneracy are then introduced to open a band gap, while tracking the evolution of the relevant modes. The resulting shift of the permittivity and permeability zero-crossings determines the frequency ranges where ENZ and MNZ responses emerge. Therefore, the prediction of Dirac cones not only enables the design of EMNZ photonic crystals but also provides a systematic strategy for engineering ENZ and MNZ behavior.

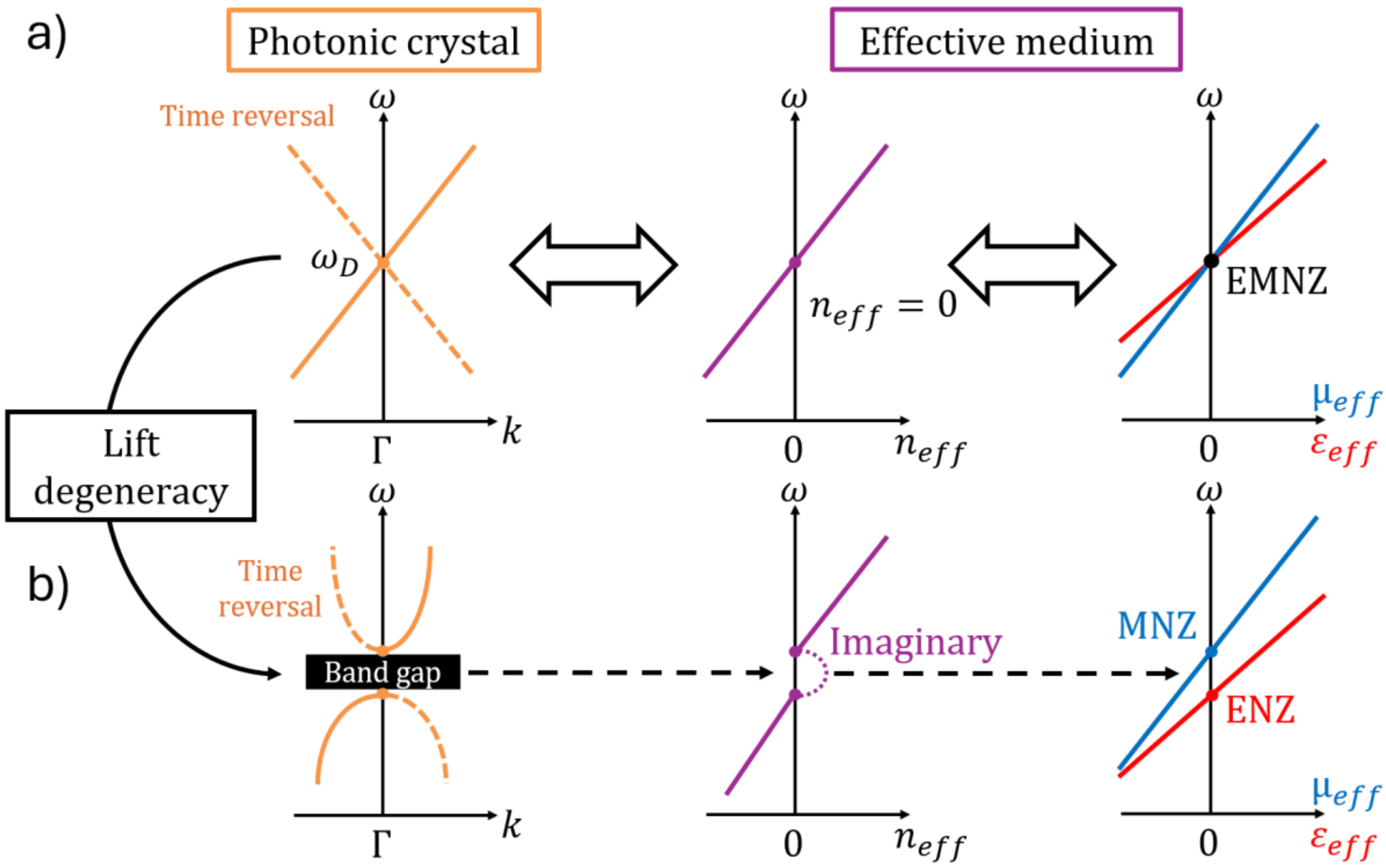

*Figure 2: a) Equivalence between Dirac cone dispersion and EMNZ behavior. A Dirac cone at the $\Gamma$-point gives rise to an EMNZ response characterized by a negative–zero–positive evolution of the effective refractive index and the simultaneous zero-crossing of the effective permittivity and permeability. Conversely, an EMNZ medium necessarily exhibits a Dirac-cone dispersion at the zero-index frequency $\omega_D$. b) Lifting the accidental degeneracy opens a band gap in the band structure. Within the gap, the refractive index becomes purely imaginary, implying opposite signs of the effective permittivity and permeability. Consequently, the zero-crossings of the two parameters split and shift to opposite sides of the band gap, giving rise to ENZ and MNZ frequency regions.*

To further establish that Dirac cones constitute the unique route toward an effective EMNZ response, we consider an alternative hypothetical scenario in which two quadratic bands with opposite curvatures become degenerate at the $\Gamma$-point. Using proof by contradiction, we show that such a degeneracy cannot satisfy the conditions required for an EMNZ medium, despite exhibiting a band crossing at zero wavevector.

When two photonic bands exhibit a quadratic degeneracy at the $\Gamma$-point, their dispersions near the degeneracy frequency $\omega_I$ can be written as $\omega_{\pm}(k) = \omega_I + \sigma_{\pm}\,|k|^2$, where the curvatures $\sigma_+$ and $\sigma_-$ have opposite signs (Figure 3). Inverting these relations gives $k_{\pm}^2(\omega) = (\omega - \omega_I)/\sigma_{\pm}$, whose derivatives with respect to frequency are $1/\sigma_{\pm}$. Because these derivatives are finite and have opposite signs, no single differentiable function $k^2(\omega)$ can represent both branches at $\omega_I$. The crystal dispersion therefore cannot be described by a $C^1$ function of frequency. In a homogeneous medium, the dispersion relation is $k^2(\omega) = \omega^2\varepsilon(\omega)\mu(\omega)/c^2$. If the effective permittivity and permeability are at least $C^1$ near $\omega_I$, then $k^2(\omega)$ is also $C^1$. For an EMNZ medium, where $\varepsilon(\omega_I) = \mu(\omega_I) = 0$, one obtains $k^2(\omega_I) = 0$ and $dk^2/d\omega|_{\omega_I} = 0$, so the dispersion remains differentiable and continuous at $\omega_I$. The quadratic degeneracy of the photonic crystal is therefore incompatible with the dispersion of a homogeneous EMNZ medium. Since the crystal requires two branches with finite slopes of opposite sign, it cannot be represented by a single differentiable effective dispersion relation.

Reproducing such behavior would require singular effective parameters (Figure 3), which is inconsistent with the definition of an EMNZ medium, in agreement with recent studies reporting similar breakdowns of the effective-medium description in such regime[61].

Consequently, the existence of an EMNZ response necessarily excludes quadratic band degeneracies of opposite curvature. Combined with the results above, this establishes Dirac-cone dispersion as the unique band-structure signature of an effective EMNZ medium.

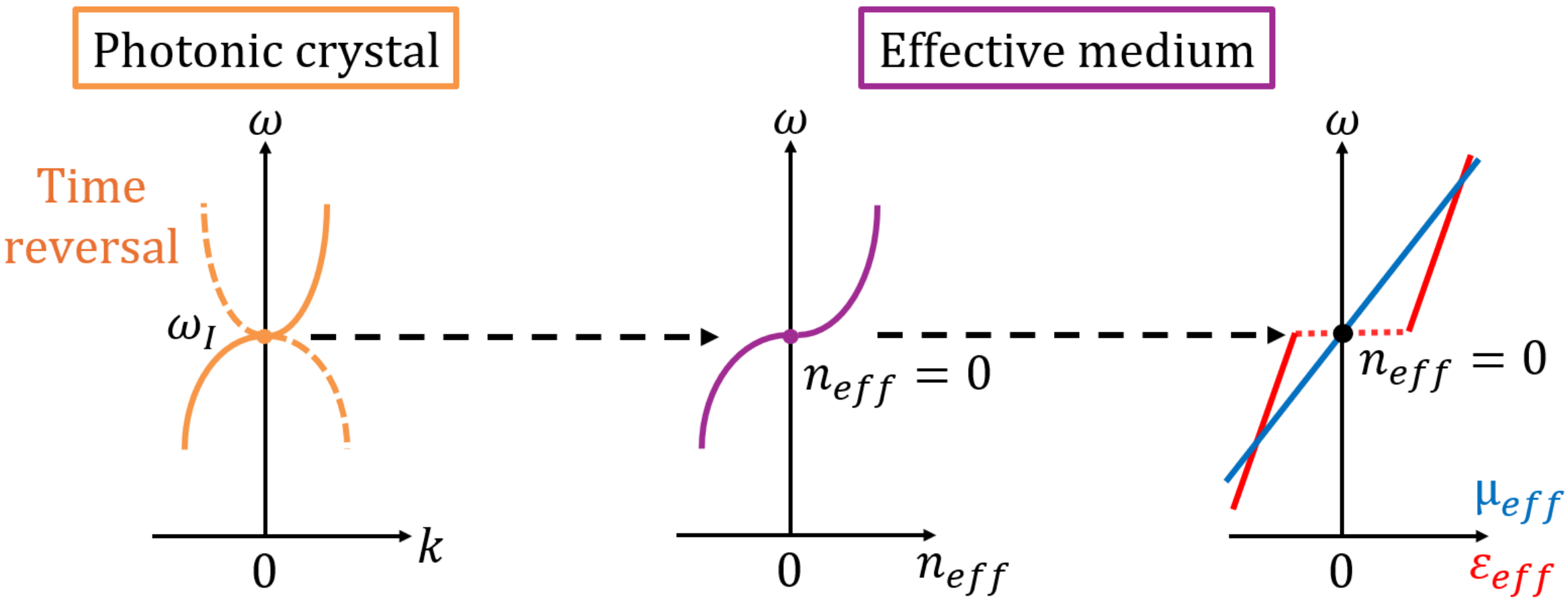

*Figure 3: Breakdown of EMNZ behavior for quadratic band degeneracies. Near the intersection of two bands with opposite curvature (orange), the photonic crystal can be formally associated with a homogeneous effective medium described by $k(\omega) = n_{eff}\omega/c$, where the effective refractive index (purple) exhibits a negative–zero–positive evolution across the degeneracy frequency $\omega_I$. However, the change in curvature at the band intersection implies that this effective description requires a discontinuity in at least one constitutive parameter (here illustrated using the permittivity $\varepsilon_{eff}$ in red), while the other parameter remains continuous (blue). The orange dotted curve corresponds to the time-reversal symmetry counterpart of the band structure.*

# Normal-incidence response of Dirac-cone photonic crystals

Previous studies have reported that some Dirac-cone photonic crystals do not exhibit all the characteristics expected from an EMNZ medium under oblique-incidence excitation[43,51–53]. In the following, we show that the EMNZ behavior is nonetheless recovered at normal incidence, indicating that the Dirac cone remains sufficient to reproduce the characteristic properties of EMNZ media at normal incidence.

To illustrate the properties of Dirac-cone photonic crystals under normal-incidence excitation, we consider two square-lattice structures supporting Dirac-like cone at the $\Gamma$-point. The first consists of dielectric cylinders with relative permittivity $\varepsilon = 12.5$ and radius $r = 0.2a_0$, arranged on a square lattice[8]. This structure exhibits a monopolar-dipolar degeneracy and a Dirac cone at the normalized frequency $f = 0.541c/a_0$ (Figure 4a). The second structure consists of hollow dielectric cylinders with relative permittivity $\varepsilon = 11.75$, outer radius $r_{out} = 0.4a_0$, and inner radius $r_{in} = 0.181a_0$[43]. In this case, a quadrupolar-dipolar degeneracy gives rise to a Dirac cone at $f = 0.6c/a_0$ (Figure 4b). While the first structure is commonly regarded as an EMNZ photonic crystal[8], Chan *et al.* argued that the second one cannot be described as an effective EMNZ medium despite exhibiting a Dirac cone at the $\Gamma$-point due to coupling to the flat-band modes under oblique-incidence excitation[43].

Nevertheless, at normal incidence and at the Dirac frequency, the incident plane wave couples directly to the $\Gamma$-point of the band structure in both photonic crystals. As shown in Figure 4c-d, the phase remains nearly uniform throughout the structures, with negligible phase accumulation from one unit cell to the next. Such a uniform phase distribution is a characteristic signature of near-zero-index propagation, since the effective wavevector approaches zero and the electromagnetic field experiences almost no spatial phase variation[1–6]. These observations are further confirmed by the effective-parameter retrievals, which show that both the effective permittivity and permeability simultaneously vanish at the Dirac frequency, demonstrating an EMNZ response in both photonic crystals (Figure 4a-b).

These results demonstrate that photonic crystals exhibiting a Dirac cone recover the characteristic EMNZ response at normal incidence. Any apparent discrepancy with previous studies[43,51–53] therefore concerns the validity of the effective-medium description under more general excitation conditions, rather than the behavior at the Dirac frequency itself.

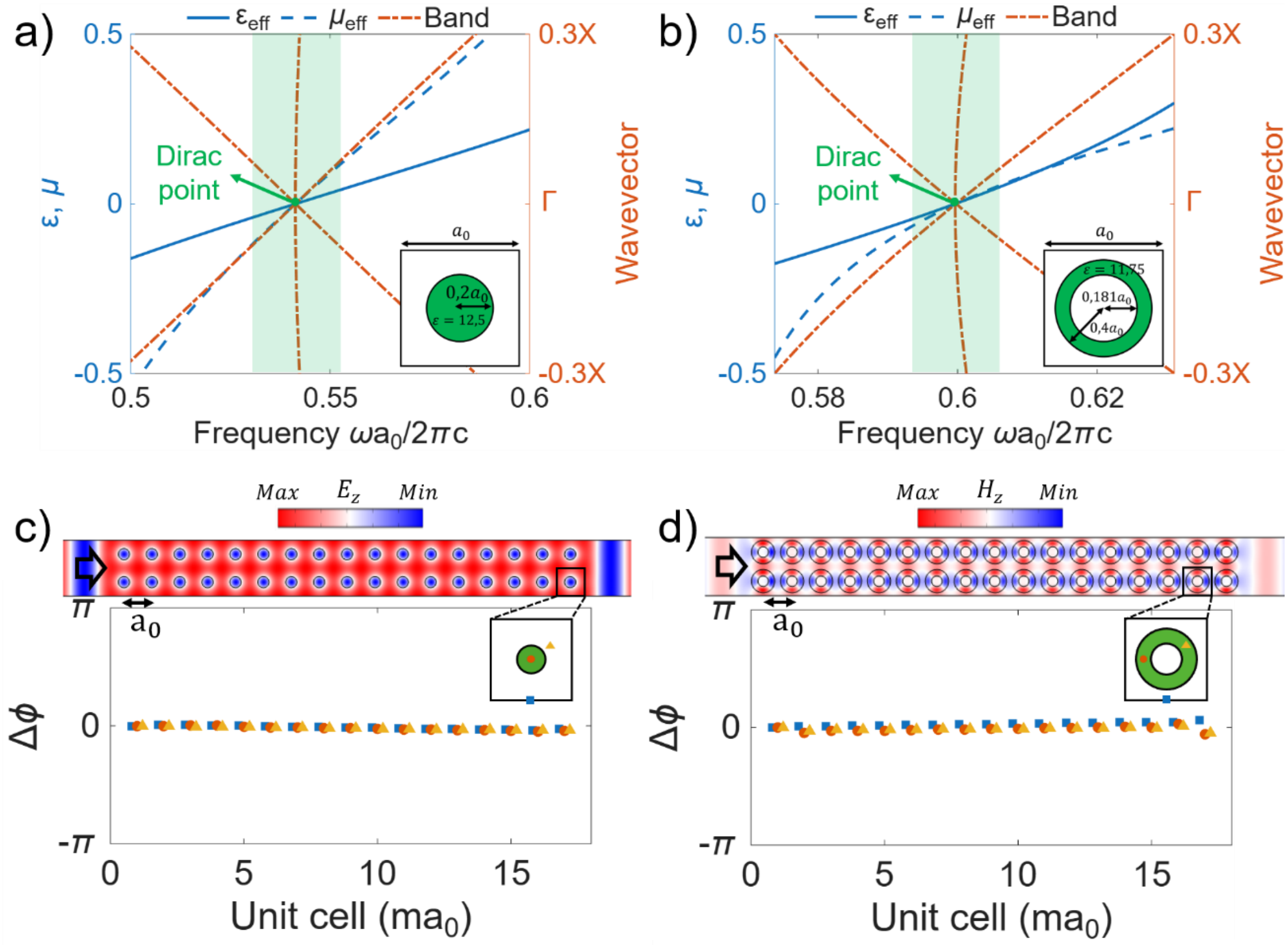

*Figure 4: (a,b) Band structures (orange dotted lines), effective parameters (effective permittivity full blue line and effective permeability dotted blue line), (c,d) field maps and phase shift Δϕ, for two photonic crystals exhibiting a Dirac cone at the Γ-point. (a,c) Square lattice of dielectric cylinders ($\varepsilon = 12.5$, $r = 0.2a_0$, TM polarization). (b,d) Square lattice of hollow cylinders ($\varepsilon = 11.75$, $r_{out} = 0.4a_0$, $r_{in} = 0.181a_0$, TE polarization) showing the corresponding band structure and effective parameters around the Dirac frequency. In both cases, the field maps and phase shift Δϕ are evaluated at Dirac point (c) $f =$*

*$0.541c/a_0$ and d) $f = 0.6c/a_0$) between three positions within the unit cell (blue square, orange circle, and yellow triangle), highlighting the spatially uniform phase evolution across the structure. The vanishing phase shift and uniform phase distribution, together with the simultaneous zero crossing of the effective permittivity and permeability, confirm the EMNZ response at the Dirac point.*

# Application to $C_{6v}$ photonic crystal

The proposed methodology (Figure 1) is now applied to the two-dimensional triangular lattice of air holes in an InGaAsP ($n = 3.26$) membrane previously reported by Contractor *et al.* to exhibit a Dirac cone at the Γ-point (Figure 5a)[48]. First, the crystal symmetry is identified as $C_{6v}$, allowing the use of Sakoda's rules to predict which modal degeneracies can lead to Dirac-cone formation[58,59]. The ratio $r/a_0$ is then varied, and the eigenmodes at the Γ-point are classified according to their irreducible representations using symmetry operations (Figure 5a). By tracking the evolution of these modes, a degeneracy between the $B_1$ and $E_2$ modes is found at $r/a_0 = 0.222$ (green line and dot on Figure 5a), consistent with previous results by Contractor *et al.*[48], satisfying Sakoda's condition for a Dirac cone. A band structure calculation (see Materials and Method) around the Γ-point confirmed the presence of the expected Dirac cone for $r/a_0 = 0.222$ (Figure 5b). Finally, effective parameter retrieval (see Materials and Method) showed that both the effective permittivity and permeability vanish at the Dirac point, demonstrating that the structure behaves as an EMNZ medium (Figure 5b). This example illustrates how the symmetry-based approach can efficiently identify geometries supporting Dirac cones and NZI behavior without relying on extensive trial and error band-structure calculations alone (See supplementary for the application on $C_{4v}$ lattice).

Furthermore, the method also captures the transition from an EMNZ regime to separate ENZ and MNZ responses. For this triangular lattice, increasing the hole radius to $r/a_0 = 0.23$ (Figure 5a, black line, blue and red dots) opens a band gap near the former Dirac point, causing the zero-crossing frequencies of the effective permittivity and permeability to split and move to opposite sides of the gap. As a result, the structure no longer exhibits simultaneous $\varepsilon_{eff} = 0$ and $\mu_{eff} = 0$, but instead supports distinct ENZ and MNZ regimes (Figure 5c). Here, ENZ response occurs at $\omega = 0.67$ and MNZ at $\omega = 0.674$.

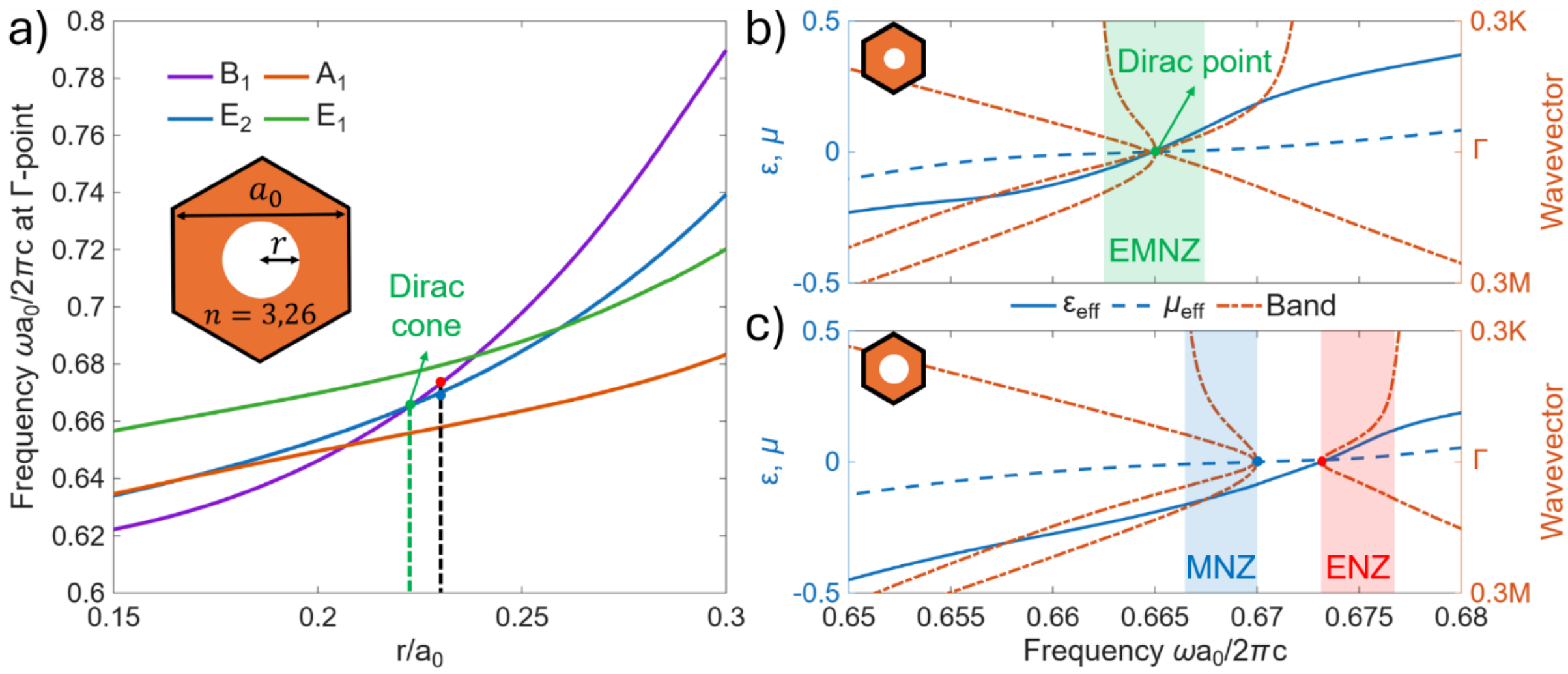

*Figure 3: a) Normalized frequency at the $\Gamma$-point for the different modes of the $C_{6v}$ group in TE polarization as a function of the ratio between the cylinder radius and the lattice period $a_0$ for a triangular lattice of air holes in InGaAsP ($n = 3.26$) slab. b) Band structure (orange dotted lines) in TE polarization and effective parameters (permittivity full blue line and permeability dotted blue line) for $r/a_0 = 0.222$ where the structure clearly shows a Dirac cone and EMNZ behavior and c) for $r/a_0 = 0.23$ where a band gap is opened and distinct ENZ and MNZ regimes appear at $\omega_{MNZ} = 0.67$ and $\omega_{ENZ} = 0.674$ respectively.*

## Application to $C_{2v}$ photonic crystal

Our methodology (Figure 1) is also applied to a $C_{2v}$ photonic crystal consisting of a rectangular array of dielectric rods ($\varepsilon = 12.5$), with lattice periods $a_y = a_0$ and $a_x = 2a_0$ (Figure 6a). By classifying the Γ-point modes according to the irreducible representations of the $C_{2v}$ group and tracking their evolution with $r/a_0$, a symmetry-allowed degeneracy between $A_1$ and $B_1$modes[58] for $r/a_0 = 0{,}303$ is identified (Figure 6a) and confirmed through band-structure calculations (Figure 6b). The resulting dispersion exhibits a semi-Dirac cone, with a linear dispersion along one direction and a quadratic dispersion along the orthogonal direction (See supplementary), providing an interesting test of the universality of symmetry-based arguments in lower-symmetry systems. Effective-parameter retrieval further reveals the emergence of an EMNZ regime at the semi-Dirac frequency, where both $\varepsilon_{eff}$ and $\mu_{eff}$ simultaneously cross zero in the $\Gamma X$ direction (Figure 6b). Indeed, the anisotropy of the semi-Dirac cone causes anisotropy in the effective refractive index, which will differ in the $\Gamma X$ and $\Gamma Y$ directions. Decreasing the radius to $r/a_0 = 0.293$ lifts the degeneracy and opens a band gap, causing the zero crossings of $\varepsilon_{eff}$ and $\mu_{eff}$ to separate on either side of the gap (Figure 6c). Consequently, the EMNZ response splits into distinct ENZ and MNZ regimes, further demonstrating how the method can predict and control the transition between these different near-zero electromagnetic behaviors through simple geometrical tuning, even in lower-symmetry systems.

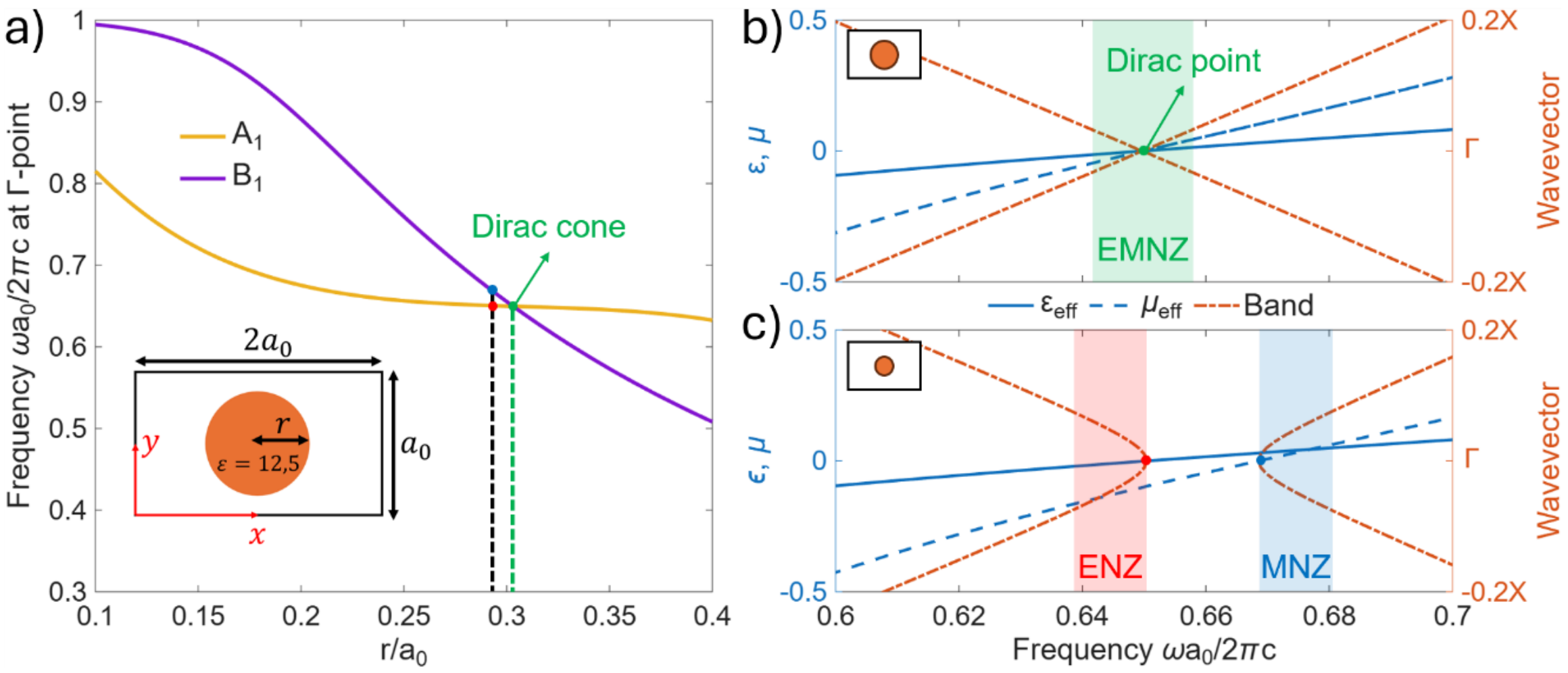

*Figure 4: a) Normalized frequency at the $\Gamma$-point for the different modes of the $C_{2v}$ group in TM polarization as a function of the ratio between the cylinder radius and the lattice period $a_0$ for a rectangular lattice ($a_y = a_x/2 = a_0$) of dielectric rods ($\varepsilon = 12.5$). b) Band structure (orange dotted line) in TE polarization and effective parameters (permittivity full blue line and permeability dotted blue line) for $r/a_0 = 0.303$ where the structure clearly shows a Dirac cone and EMNZ behavior and c) for $r/a_0 = 0.293$ where a band gap is opened and distinct ENZ and MNZ regimes appear at $\omega_{ENZ} = 0.65$ and $\omega_{MNZ} = 0.67$ respectively.*

# 3 Conclusion

In this work, we established a universal and predictive framework for the design of near-zero-index photonic crystals. We demonstrated that the presence of a Dirac cone at the $\Gamma$-point is both a necessary and sufficient condition for obtaining an effective EMNZ response under normal-incidence excitation. By proving the equivalence between Dirac cone dispersion and EMNZ behavior, and by showing that alternative quadratic degeneracies cannot satisfy the requirements of an EMNZ medium, we identified Dirac cones as the unique band-structure signature of effective EMNZ photonic crystals.

Building upon this result, we reformulated the design of NZI photonic crystals as a symmetry problem. Using mode classification based on group-theory, geometrical conditions leading to accidental degeneracies and Dirac cone formation can be predicted directly from the modal symmetries, providing a practical alternative to extensive numerical parameter sweeps[58,59]. Furthermore, we showed that controlled lifting of these degeneracies offers a systematic route toward the realization of ENZ and MNZ photonic crystals through the opening of a band gap and the separation of the permittivity and permeability zero-crossings.

The generality of the framework was illustrated through its application to photonic crystals belonging to the $C_{2v}$, $C_{4v}$ and $C_{6v}$ symmetry groups, demonstrating its ability to predict and engineer EMNZ, ENZ, and MNZ responses in distinct photonic crystal platforms. Beyond the specific examples considered here, the proposed methodology provides a unified approach for the analysis and design of near-zero-index photonic structures and offers new opportunities for the systematic exploration of NZI phenomena in periodic media.

# 4 Method

All simulations were performed using COMSOL Multiphysics 6.2 in a two-dimensional configuration, employing the *"Electromagnetic Waves, Frequency Domain"* module. Band structures were computed using the *"Eigenfrequency"* study, while wave propagation through the photonic crystals was simulated using the *"Frequency Domain"* study (see Supplementary Information for details of the numerical implementation). Although the present study is restricted to two-dimensional photonic crystals, the underlying design framework can be extended to three-dimensional structures. In this case, the thickness becomes an additional geometrical parameter that can be adjusted to achieve the degeneracy of the modes of interest.

The effective parameters were retrieved using two complementary approaches. The first is a boundary effective-medium method[49,50], in which the electromagnetic fields are averaged along the boundaries of the unit cell and the effective permittivity and permeability are extracted using Maxwell's constitutive relations. The second method is based on the phase-shift analysis[47,65], where the phase accumulation during propagation through the crystal is related to the real part of the effective refractive index, while the attenuation of the field amplitude provides access to its imaginary part (see Supplementary Information for the formalism of effective index calculations).

# Supporting Information

Application of the method to a $C_{4v}$-symmetric photonic crystal, illustration of the semi-Dirac cone, and details of the COMSOL codes.

# Author Contributions

A.D. developed the framework, derived the analytical modeling and performed the numerical simulations under the guidance and supervision of M.L., A.D. and M.L. performed the underlying analysis of the results. A.D. wrote the manuscript with input from M.L.. All the authors subsequently took part in the revision process and approved the final copy of the manuscript.

# Funding

A.D is a Research Fellow of the Fonds de la Recherche Scientifique – FNRS. M.L. is a Research Associate of the Fonds de la Recherche Scientifique – FNRS.

# Acknowledgements

This research used resources of the "Plateforme Technologique de Calcul Intensif (PTCI)" (http://www.ptci.unamur.be) located at the University of Namur, Belgium, which is supported by the FNRS-FRFC, the Walloon Region, and the University of Namur (Conventions No.

2.5020.11, GEQ U.G006.15, 1610468, RW/GEQ2016 et U.G011.22). The PTCI is member of the "Consortium des Équipements de Calcul Intensif (CÉCI)" (http://www.ceci-hpc.be). Generative AI tools were used during the preparation of this manuscript to assist with text drafting, language editing, and for translation purposes only. All AI-generated content was subsequently reviewed, modified, and validated by the authors, who take full responsibility for the final content of the manuscript.

# References

(1) Kinsey, N.; DeVault, C.; Boltasseva, A.; Shalaev, V. M. Near-Zero-Index Materials for Photonics. *Nat. Rev. Mater.* **2019**, *4* (12), 742–760. https://doi.org/10.1038/s41578-019-0133-0.

(2) Liberal, I.; Engheta, N. Near-Zero Refractive Index Photonics. *Nat. Photonics* **2017**, *11* (3), 149–158. https://doi.org/10.1038/nphoton.2017.13.

(3) Vertchenko, L.; Nikitin, M.; Lavrinenko, A. Near-Zero-Index Platform in Photonics: Tutorial. *JOSA B* **2023**, *40* (6), 1467–1482. https://doi.org/10.1364/JOSAB.489055.

(4) Vulis, D. I.; Reshef, O.; Camayd-Muñoz, P.; Mazur, E. Manipulating the Flow of Light Using Dirac-Cone Zero-Index Metamaterials. *Rep. Prog. Phys.* **2018**, *82* (1), 012001. https://doi.org/10.1088/1361-6633/aad3e5.

(5) Lobet, M.; Kinsey, N.; Liberal, I.; Caglayan, H.; Huidobro, P. A.; Galiffi, E.; Mejía-Salazar, J. R.; Palermo, G.; Jacob, Z.; Maccaferri, N. New Horizons in Near-Zero Refractive Index Photonics and Hyperbolic Metamaterials. *ACS Photonics* **2023**, *10* (11), 3805–3820. https://doi.org/10.1021/acsphotonics.3c00747.

(6) Li, Y.; Chan, C. T.; Mazur, E. Dirac-like Cone-Based Electromagnetic Zero-Index Metamaterials. *Light Sci. Appl.* **2021**, *10* (1), 203. https://doi.org/10.1038/s41377-021-00642-2.

(7) Islam, S.; Faruque, M.; Islam, M. A Near Zero Refractive Index Metamaterial for Electromagnetic Invisibility Cloaking Operation. *Materials* **2015**, *8* (8), 4790–4804. https://doi.org/10.3390/ma8084790.

(8) Huang, X.; Lai, Y.; Hang, Z. H.; Zheng, H.; Chan, C. T. Dirac Cones Induced by Accidental Degeneracy in Photonic Crystals and Zero-Refractive-Index Materials. *Nat. Mater.* **2011**, *10* (8), 582–586. https://doi.org/10.1038/nmat3030.

(9) Silveirinha, M.; Engheta, N. Tunneling of Electromagnetic Energy through Subwavelength Channels and Bends Using ε -Near-Zero Materials. *Phys. Rev. Lett.* **2006**, *97* (15), 157403. https://doi.org/10.1103/PhysRevLett.97.157403.

(10) Alù, A.; Silveirinha, M. G.; Salandrino, A.; Engheta, N. Epsilon-near-Zero Metamaterials and Electromagnetic Sources: Tailoring the Radiation Phase Pattern. *Phys. Rev. B* **2007**, *75* (15), 155410. https://doi.org/10.1103/PhysRevB.75.155410.

(11) Mello, O.; Li, Y.; Camayd-Muñoz, S. A.; DeVault, C.; Lobet, M.; Tang, H.; Lonçar, M.; Mazur, E. Extended Many-Body Superradiance in Diamond Epsilon near-Zero Metamaterials. *Appl. Phys. Lett.* **2022**, *120* (6), 061105. https://doi.org/10.1063/5.0062869.

(12) Mello, O.; Vertchenko, L.; Nelson, S.; Debacq, A.; Guney, D.; Mazur, E.; Lobet, M. Long-Range Quantum Entanglement in Dielectric Mu-near-Zero Metamaterials. *Light Sci. Appl.* **2025**, *14* (1), 300. https://doi.org/10.1038/s41377-025-01994-9.

(13) Ren, J.; Zhu, S.; Wang, Z. D. Attaining Nearly Ideal Dicke Superradiance in Expanded Spatial Domains. *Phys. Rev. Appl.* **2024**, *21* (4), 044053. https://doi.org/10.1103/PhysRevApplied.21.044053.

(14) Fleury, R.; Alù, A. Enhanced Superradiance in Epsilon-near-Zero Plasmonic Channels. *Phys. Rev. B* **2013**, *87* (20), 201101. https://doi.org/10.1103/PhysRevB.87.201101.

(15) Alù, A.; Engheta, N. Emission Enhancement in a Plasmonic Waveguide at Cut-Off. *Materials* **2011**, *4* (1), 141–152. https://doi.org/10.3390/ma4010141.
(16) Sokhoyan, R.; Atwater, H. A. Quantum Optical Properties of a Dipole Emitter Coupled to an Ɛ-near-Zero Nanoscale Waveguide. *Opt. Express* **2013**, *21* (26), 32279. https://doi.org/10.1364/OE.21.032279.
(17) Özgün, E.; Ozbay, E.; Caglayan, H. Tunable Zero-Index Photonic Crystal Waveguide for Two-Qubit Entanglement Detection. *ACS Photonics* **2016**, *3* (11), 2129–2133. https://doi.org/10.1021/acsphotonics.6b00576.
(18) Alam, M. Z.; De Leon, I.; Boyd, R. W. Large Optical Nonlinearity of Indium Tin Oxide in Its Epsilon-near-Zero Region. *Science* **2016**, *352* (6287), 795–797. https://doi.org/10.1126/science.aae0330.
(19) Caspani, L.; Kaipurath, R. P. M.; Clerici, M.; Ferrera, M.; Roger, T.; Kim, J.; Kinsey, N.; Pietrzyk, M.; Di Falco, A.; Shalaev, V. M.; Boltasseva, A.; Faccio, D. Enhanced Nonlinear Refractive Index in ε -Near-Zero Materials. *Phys. Rev. Lett.* **2016**, *116* (23), 233901. https://doi.org/10.1103/PhysRevLett.116.233901.
(20) Kinsey, N.; Khurgin, J. Nonlinear Epsilon-near-Zero Materials Explained: Opinion. *Opt. Mater. Express* **2019**, *9* (7), 2793. https://doi.org/10.1364/OME.9.002793.
(21) Reshef, O.; De Leon, I.; Alam, M. Z.; Boyd, R. W. Nonlinear Optical Effects in Epsilon-near-Zero Media. *Nat. Rev. Mater.* **2019**, *4* (8), 535–551. https://doi.org/10.1038/s41578-019-0120-5.
(22) Javani, M. H.; Stockman, M. I. Real and Imaginary Properties of Epsilon-Near-Zero Materials. *Phys. Rev. Lett.* **2016**, *117* (10), 107404. https://doi.org/10.1103/PhysRevLett.117.107404.
(23) Lobet, M.; Liberal, I.; Knall, E. N.; Alam, M. Z.; Reshef, O.; Boyd, R. W.; Engheta, N.; Mazur, E. Fundamental Radiative Processes in Near-Zero-Index Media of Various Dimensionalities. *ACS Photonics* **2020**, *7* (8), 1965–1970. https://doi.org/10.1021/acsphotonics.0c00782.
(24) Lobet, M.; Liberal, I.; Vertchenko, L.; Lavrinenko, A. V.; Engheta, N.; Mazur, E. Momentum Considerations inside Near-Zero Index Materials. *Light Sci. Appl.* **2022**, *11* (1), 110. https://doi.org/10.1038/s41377-022-00790-z.
(25) Ziolkowski, R. W. Propagation in and Scattering from a Matched Metamaterial Having a Zero Index of Refraction. *Phys. Rev. E* **2004**, *70* (4), 046608. https://doi.org/10.1103/PhysRevE.70.046608.
(26) Moitra, P.; Yang, Y.; Anderson, Z.; Kravchenko, I. I.; Briggs, D. P.; Valentine, J. Realization of an All-Dielectric Zero-Index Optical Metamaterial. *Nat. Photonics* **2013**, *7* (10), 791–795. https://doi.org/10.1038/nphoton.2013.214.
(27) Tang, H.; DeVault, C.; Camayd-Muñoz, S. A.; Liu, Y.; Jia, D.; Du, F.; Mello, O.; Vulis, D. I.; Li, Y.; Mazur, E. Low-Loss Zero-Index Materials. *Nano Lett.* **2021**, *21* (2), 914–920. https://doi.org/10.1021/acs.nanolett.0c03575.
(28) Li, Y.; Kita, S.; Muñoz, P.; Reshef, O.; Vulis, D. I.; Yin, M.; Lončar, M.; Mazur, E. On-Chip Zero-Index Metamaterials. *Nat. Photonics* **2015**, *9* (11), 738–742. https://doi.org/10.1038/nphoton.2015.198.
(29) Silveirinha, M. G.; Engheta, N. Theory of Supercoupling, Squeezing Wave Energy, and Field Confinement in Narrow Channels and Tight Bends Using ε near-Zero Metamaterials. *Phys. Rev. B* **2007**, *76* (24), 245109. https://doi.org/10.1103/PhysRevB.76.245109.
(30) Alù, A.; Silveirinha, M. G.; Engheta, N. Transmission-Line Analysis of ε -near-Zero–Filled Narrow Channels. *Phys. Rev. E* **2008**, *78* (1), 016604. https://doi.org/10.1103/PhysRevE.78.016604.
(31) Marcos, J. S.; Silveirinha, M. G.; Engheta, N. *M* -near-Zero Supercoupling. *Phys. Rev. B* **2015**, *91* (19), 195112. https://doi.org/10.1103/PhysRevB.91.195112.
(32) Mahmoud, A. M.; Engheta, N. Wave–Matter Interactions in Epsilon-and-Mu-near-Zero Structures. *Nat. Commun.* **2014**, *5* (1), 5638. https://doi.org/10.1038/ncomms6638.

(33) Li, Y.; Argyropoulos, C. Controlling Collective Spontaneous Emission with Plasmonic Waveguides. *Opt. Express* **2016**, *24* (23), 26696. https://doi.org/10.1364/OE.24.026696.
(34) Alù, A.; Engheta, N. Boosting Molecular Fluorescence with a Plasmonic Nanolauncher. *Phys. Rev. Lett.* **2009**, *103* (4), 043902. https://doi.org/10.1103/PhysRevLett.103.043902.
(35) So, J.-K.; Yuan, G. H.; Soci, C.; Zheludev, N. I. Enhancement of Luminescence of Quantum Emitters in Epsilon-near-Zero Waveguides. *Appl. Phys. Lett.* **2020**, *117* (18), 181104. https://doi.org/10.1063/5.0018488.
(36) Edwards, B.; Alù, A.; Young, M. E.; Silveirinha, M.; Engheta, N. Experimental Verification of Epsilon-Near-Zero Metamaterial Coupling and Energy Squeezing Using a Microwave Waveguide. *Phys. Rev. Lett.* **2008**, *100* (3), 033903. https://doi.org/10.1103/PhysRevLett.100.033903.
(37) Vesseur, E. J. R.; Coenen, T.; Caglayan, H.; Engheta, N.; Polman, A. Experimental Verification of n = 0 Structures for Visible Light. *Phys. Rev. Lett.* **2013**, *110* (1), 013902. https://doi.org/10.1103/PhysRevLett.110.013902.
(38) Pendry, J. B.; Holden, A. J.; Robbins, D. J.; Stewart, W. J. Magnetism from Conductors and Enhanced Nonlinear Phenomena. *IEEE Trans. Microw. Theory Tech.* **1999**, *47* (11), 2075–2084. https://doi.org/10.1109/22.798002.
(39) Belov, P. A.; Slobozhanyuk, A. P.; Filonov, D. S.; Yagupov, I. V.; Kapitanova, P. V.; Simovski, C. R.; Lapine, M.; Kivshar, Y. S. Broadband Isotropic μ-near-Zero Metamaterials. *Appl. Phys. Lett.* **2013**, *103* (21), 211903. https://doi.org/10.1063/1.4832056.
(40) Liberal, I.; Mahmoud, A. M.; Li, Y.; Edwards, B.; Engheta, N. Photonic Doping of Epsilon-near-Zero Media. *Science* **2017**, *355* (6329), 1058–1062. https://doi.org/10.1126/science.aal2672.
(41) Shelby, R. A.; Smith, D. R.; Schultz, S. Experimental Verification of a Negative Index of Refraction. *Science* **2001**, *292* (5514), 77–79. https://doi.org/10.1126/science.1058847.
(42) Dolling, G.; Enkrich, C.; Wegener, M.; Soukoulis, C. M.; Linden, S. Simultaneous Negative Phase and Group Velocity of Light in a Metamaterial. *Science* **2006**, *312* (5775), 892–894. https://doi.org/10.1126/science.1126021.
(43) Chan, C. T.; Hang, Z. H.; Huang, X. Dirac Dispersion in Two-Dimensional Photonic Crystals. *Adv. Optoelectron.* **2012**, *2012*, 1–11. https://doi.org/10.1155/2012/313984.
(44) Ochiai, T.; Onoda, M. Photonic Analog of Graphene Model and Its Extension: Dirac Cone, Symmetry, and Edge States. *Phys. Rev. B* **2009**, *80* (15), 155103. https://doi.org/10.1103/PhysRevB.80.155103.
(45) Castro Neto, A. H.; Guinea, F.; Peres, N. M. R.; Novoselov, K. S.; Geim, A. K. The Electronic Properties of Graphene. *Rev. Mod. Phys.* **2009**, *81* (1), 109–162. https://doi.org/10.1103/RevModPhys.81.109.
(46) Dong, T.; Liang, J.; Camayd-Muñoz, S.; Liu, Y.; Tang, H.; Kita, S.; Chen, P.; Wu, X.; Chu, W.; Mazur, E.; Li, Y. Ultra-Low-Loss on-Chip Zero-Index Materials. *Light Sci. Appl.* **2021**, *10* (1), 10. https://doi.org/10.1038/s41377-020-00436-y.
(47) Minkov, M. Zero-Index Bound States in the Continuum. *Phys. Rev. Lett.* **2018**, *121* (26). https://doi.org/10.1103/PhysRevLett.121.263901.
(48) Contractor, R.; Noh, W.; Redjem, W.; Qarony, W.; Martin, E.; Dhuey, S.; Schwartzberg, A.; Kanté, B. Scalable Single-Mode Surface-Emitting Laser via Open-Dirac Singularities. *Nature* **2022**, *608* (7924), 692–698. https://doi.org/10.1038/s41586-022-05021-4.
(49) Vertchenko, L.; DeVault, C.; Malureanu, R.; Mazur, E.; Lavrinenko, A. Near-Zero Index Photonic Crystals with Directive Bound States in the Continuum. *Laser Photonics Rev.* **2021**, *15* (7), 2000559. https://doi.org/10.1002/lpor.202000559.
(50) Wu, Y. A Semi-Dirac Point and an Electromagnetic Topological Transition in a Dielectric Photonic Crystal. *Opt. Express* **2014**, *22* (2), 1906. https://doi.org/10.1364/OE.22.001906.

(51) Gao, H.; Zhou, Y. S.; Zheng, Z. Y.; Chen, S. J.; Dong, J. J. Is the Photonic Crystal with a Dirac Cone at Its Γ Point a Real Zero-Index Material? *Appl. Phys. B* **2017**, *123* (5), 165. https://doi.org/10.1007/s00340-017-6738-3.
(52) Wei, G.-G.; Miao, C.; Huang, H.-C.; Gao, H. Zero Refractive Index Properties of Two-Dimensional Photonic Crystals with Dirac Cones. *Chin. Phys. Lett.* **2019**, *36* (3), 034203. https://doi.org/10.1088/0256-307X/36/3/034203.
(53) Ashraf, M. W.; Faryad, M. On the Mapping of Dirac-like Cone Dispersion in Dielectric Photonic Crystals to an Effective Zero-Index Medium. *J. Opt. Soc. Am. B* **2016**, *33* (6), 1008. https://doi.org/10.1364/JOSAB.33.001008.
(54) Zhang, P.; Fietz, C.; Tassin, P.; Koschny, T.; Soukoulis, C. M. Numerical Investigation of the Flat Band Bloch Modes in a 2D Photonic Crystal with Dirac Cones. *Opt. Express* **2015**, *23* (8), 10444. https://doi.org/10.1364/OE.23.010444.
(55) Wu, Y.; Li, J.; Zhang, Z.-Q.; Chan, C. T. Effective Medium Theory for Magnetodielectric Composites: Beyond the Long-Wavelength Limit. *Phys. Rev. B* **2006**, *74* (8), 085111. https://doi.org/10.1103/PhysRevB.74.085111.
(56) Śmigaj, W.; Gralak, B. Validity of the Effective-Medium Approximation of Photonic Crystals. *Phys. Rev. B* **2008**, *77* (23), 235445. https://doi.org/10.1103/PhysRevB.77.235445.
(57) Markel, V. A.; Tsukerman, I. Applicability of Effective Medium Description to Photonic Crystals in Higher Bands: Theory and Numerical Analysis. *Phys. Rev. B* **2016**, *93* (22), 224202. https://doi.org/10.1103/PhysRevB.93.224202.
(58) Sakoda, K. Proof of the Universality of Mode Symmetries in Creating Photonic Dirac Cones. *Opt. Express* **2012**, *20* (22), 25181. https://doi.org/10.1364/OE.20.025181.
(59) Sakoda, K. Universality of Mode Symmetries in Creating Photonic Dirac Cones. *JOSA B* **2012**, *29* (10), 2770–2778. https://doi.org/10.1364/JOSAB.29.002770.
(60) Li, Y.; Wu, Y.; Chen, X.; Mei, J. Selection Rule for Dirac-like Points in Two-Dimensional Dielectric Photonic Crystals. *Opt. Express* **2013**, *21* (6), 7699. https://doi.org/10.1364/OE.21.007699.
(61) Vallar, E.; Madera, C. R.; Santagiustina, M.; Semenova, E.; Huck, A.; Lavrinenko, A. A General Method of Wavelength-Tuning of Dirac-like Cones in Photonic Crystals. May 14, 2026. https://doi.org/10.1364/opticaopen.32262279.
(62) Wang, L.-G.; Wang, Z.-G.; Zhang, J.-X.; Zhu, S.-Y. Realization of Dirac Point with Double Cones in Optics. *Opt. Lett.* **2009**, *34* (10), 1510. https://doi.org/10.1364/OL.34.001510.
(63) Alù, A.; Yaghjian, A. D.; Shore, R. A.; Silveirinha, M. G. Causality Relations in the Homogenization of Metamaterials. *Phys. Rev. B* **2011**, *84* (5), 054305. https://doi.org/10.1103/PhysRevB.84.054305.
(64) Milonni, P. W.; Maclay, G. J. Quantized-Field Description of Light in Negative-Index Media. *Opt. Commun.* **2003**, *228* (1–3), 161–165. https://doi.org/10.1016/j.optcom.2003.09.080.
(65) Andryieuski, A.; Malureanu, R.; Lavrinenko, A. V. Wave Propagation Retrieval Method for Metamaterials: Unambiguous Restoration of Effective Parameters. *Phys. Rev. B* **2009**, *80* (19), 193101. https://doi.org/10.1103/PhysRevB.80.193101.